\documentclass[aps,prb,showpacs,twocolumn,floats]{revtex4}
\usepackage{amssymb}

\usepackage{graphicx,psfig}
\usepackage{dcolumn}
\usepackage{bm}

\input{tcilatex}

\begin{document}

\title{Universal mechanism of spin relaxation in solids}
\author{E. M. Chudnovsky,$^1$ D. A. Garanin,$^2$ and R. Schilling$^2$}
\affiliation{ \mbox{$^1$Physics Department, Lehman College, City
University of New York,} \\ \mbox{250 Bedford
Park Boulevard West, Bronx, New York 10468-1589, U.S.A.} \\
\mbox{$^2$Institut f\"ur Physik, Johannes-Gutenberg-Universit\"at,
 D-55099 Mainz, Germany}}
\date{9 March 2005}

\begin{abstract}
We consider relaxation of a rigid spin cluster in an elastic medium in the
presence of the magnetic field. Universal simple expression for spin-phonon
matrix elements due to local rotations of the lattice is derived. The
equivalence of the lattice frame and the laboratory frame approaches is
established. For spin Hamiltonians with strong uniaxial anisotropy the field
dependence of the transition rates due to rotations is analytically
calculated and its universality is demonstrated. The role of time reversal
symmetry in spin-phonon transitions has been elucidated. The theory provides
lower bound on the decoherence of any spin-based solid-state qubit.
\end{abstract}
 \maketitle


\section{Introduction}

Understanding spin-lattice interactions has fundamental importance for
applications of magnetic phenomena, such as, e.g., magnetic relaxation,
magnetic resonance, and decoherence of spin-based qubits. The studies of
spin-lattice interactions are almost as old as the quantum theory of solids.
\cite{wal32,heitel34,kro39,vle40,orb61} Van Vleck \cite{vle40} was among the
first who attempted to compute the rates of phonon-induced transitions (in
titanium and chromalum) from the first principles, that is, considering
atomic wave functions in the crystal field and spin-orbit interaction. These
calculations are very involved, and they are hardly possible for more
complicated materials such as magnetic molecules or clusters, each cluster
containing several magnetic atoms and tens or hundreds of nonmagnetic atoms.

On the other hand, for most materials the orbital moment of electrons is
quenched by a strong crystal field and it does not contribute to the
magnetic properties. The latter are due to the spin $\mathbf{S}$ that is
relatively weakly coupled to the orbital moment via the spin-orbit
interaction and thus it feels the crystal field indirectly$\mathbf{.}$ While
microscopic calculation of the crystal-field Hamiltonian for $\mathbf{S}$ is
difficult, one can use an alternative approach\cite{orb61,abrble70} and
start with a phenomenological expression containing all permitted by
symmetry combinations of spin $\mathbf{S}$. Some examples of the
crystal-field Hamiltonian that results in the uniaxial, biaxial, and cubic
magnetic anisotropy of the cluster are given below:
\begin{eqnarray}
\hat{H}_{A} &=&-D(\mathbf{S}\cdot \mathbf{e}^{(3)})^{2}  \label{uniaxial} \\
\hat{H}_{A} &=&-D(\mathbf{S}\cdot \mathbf{e}^{(3)})^{2}+E[(\mathbf{S}\cdot
\mathbf{e}^{(1)})^{2}-(\mathbf{S}\cdot \mathbf{e}^{(2)})^{2}]
\label{biaxial} \\
\hat{H}_{A} &=&C\sum_{\alpha \neq \beta }(\mathbf{S}\cdot \mathbf{e}%
^{(\alpha )})^{2}(\mathbf{S}\cdot \mathbf{e}^{(\beta )})^{2},  \label{cubic}
\end{eqnarray}
where $\mathbf{e}^{(\alpha )}$ with $\alpha =1,2,3$ are the unit vectors of
the coordinate frame that is rigidly coupled with the symmetry axes of the
cluster (in the following it will be called ``lattice frame''). The
advantage of this approach is that the form of $\hat{H}_{A}$ is uniquely
determined by symmetry, while the phenomenological constants ($D$, $E$, $C$,
etc.) can be easily obtained from experiment. The expressions for $\hat{H}%
_{A}$ above, as well as all other physically acceptable forms of $\hat{H}%
_{A},$ possess a full rotational invariance, i.e., the form of
$\hat{H}_{A}$ does not depend on the orientation of laboratory
coordinate axes. The full rotational invariance implies
conservation of the total angular momentum (spin + lattice). Eqs.\
(\ref{uniaxial})--(\ref{cubic}) are also applicable to magnetic
molecules or clusters, $\mathbf{S}$ being the total spin of the
cluster formed by a strong exchange interaction between the
individual magnetic atoms.

Lattice vibrations couple to the spin by modifying coefficients in
Eqs.\ (\ref
{uniaxial})--(\ref{cubic}), changing directions of the lattice vectors $%
\mathbf{e}^{(\alpha )},$ as well as by lowering the symmetry of the crystal
field. Due to translational invariance, the phonon displacement field $%
\mathbf{u}$ enters the Hamiltonian only through its spatial derivatives $%
\partial u_{i}/\partial x_{j}$ or their symmetric and antisymmetric
combinations. Again, one can write down a phenomenological expression for
the spin-phonon Hamiltonian $\hat{H}_{\mathrm{s-ph}}$ that contains all
terms permitted by the symmetry of a particular material.\cite{calcal65pr}
However the general form of $\hat{H}_{\mathrm{s-ph}}$ usually contains too
many different coefficients that are comparable with each other and cannot
be measured independently.

The problem of spin-lattice relaxation can be simplified if one notices that
longitudinal phonons have a larger sound velocity than the transverse
phonons. Since the rate of one-phonon processes (emission and absorption of
a phonon) is inversely proportional to the fifth power of the sound
velocity, procecces involving longitudinal phonons can be neglected. The
same is valid for multiphonon processes such as the Raman process, because
their rates contain even higher powers of the sound velocity.

The terms of $\hat{H}_{\mathrm{s-ph}}$ due to transverse phonons can be
split into two groups. The first group describes distorsions of the lattice
cell due to transverse phonons whereas the second group of terms describes
local rotations of the lattice without distortion of the crystal environment
of magnetic atoms. Whereas the first group of terms contains
phenomenological coupling coefficients, the terms of $\hat{H}_{\mathrm{s-ph}%
} $ due to local rotations are \emph{parameter free} and are
defined solely by the form of $\hat{H}_{A}.$ The significance of
the latter was noticed in the past,\cite
{mel72prl,bonmel76prb,mel79prb,fed75prb,fedmel76prb,fed77prb,dohful75,wanlue77prb}
and different kinds of magnetoelastic problems have been
considered. However early applications of the theory have not
included spin-lattice relaxation. Much later the relaxation
between the adjacent spin levels of the spin Hamiltonian of Eq.\
(\ref{uniaxial}) due to the parameter-free spin-lattice
interaction arising from tilting of $\mathbf{e}^{(3)}$ by
transverse phonons was considered in Ref.\ \onlinecite{garchu97}.

In general, processes due to the distortion of the lattice and those due to
the local rotation of the lattice should result in comparable relaxation
rates. Even in this case, the latter are of a fundamental importance because
they provide a parameter-free lower bound on the decoherence of any
spin-based qubit. In the case of a magnetic molecule or a cluster in a
solid, if the cluster is more rigid than its environment, it resists any
distortions due to long-wave deformations of the solid. That is, $\mathbf{S}$
interacts only with the long-wave deformations of the crystal lattice that
rotate the local frame $(\mathbf{e}^{(1)},\mathbf{e}^{(2)},\mathbf{e}^{(3)})$
as a whole, so that the spin-phonon interaction can be obtained from $\hat{H}%
_{A}$ without any phenomenological parameters. The corresponding
parameter-free description of spin-lattice relaxation becomes \emph{exact}
in this case.

It has been noticed \cite{chu94prl,chumar02prb,chu04prl,chugar04prl} that
for an arbitrary spin Hamiltonian the parameter-free spin-phonon transition
rates can be conveniently computed by switching to the lattice frame where
the form of $\hat{H}_{A}$ is preserved, while the spin-phonon interaction is
of a kinematic origin and it has a universal form that is independent from $%
\hat{H}_{A}.$ With this new method, one could easily calculate the
relaxation rates between the tunnel-split states of $\hat{H}_{A},$ while it
was unclear how this problem could be solved by conventional methods in the
laboratory frame. A striking feature of the lattice-frame approach is that
both the calculation and the final result for the relaxation rate are
universal and insensitive to the detailed form of $\hat{H}_{A}.$ One obtains
the relaxation rate that is expressed via the quantities that can be
directly measured in experiment, such as the tunnel splitting $\Delta .$

The aim of the present work is to investigate the relation between the
lattice-frame and laboratory-frame approaches in more detail. We will show
that also in the laboratory frame one can formulate a new method of
calculating parameter-free spin-phonon rates that is similar to the
lattice-frame approach and leads to the same results. We will extend the
theory by taking into account the magnetic field, including the case in
which the tunnel splitting is solely due to the magnetic field.

The body of this paper is organized as follows. In Sec.\
\ref{Sec-SpinLattice} we derive exact expressions for the
spin-phonon interaction induced by rotations in the laboratory
(Sec.\ \ref{Sec-HSPh}) and lattice (Sec.\ \ref {Sec-HSPhLatt})
frames. In Sec.\ \ref{Sec-ME} a simple universal formula for the
spin-phonon matrix element is obtained for an arbitrary spin
Hamiltonian in the presence of the magnetic field. Equivalence of
the laboratory and lattice frame treatments of the spin-phonon
interaction is demonstrated. \emph{The advantage of our method is
that it requires only the knowledge of the matrix elements of the
operator }$\mathbf{S}$\emph{\ between the eigenstates of the
Hamiltonian}, in contrast with the traditional method that
requires the knowledge of a matrix element of a model-dependent
product of the components of $\mathbf{S}.$ In Sec.\
\ref{Sec-MatrElem} spin matrix elements for transitions between
spin states split by the crystal field or by the magnetic field
are calculated analytically for Hamiltonians that are dominated by
the uniaxial anisotropy. We show that these matrix elements are
parameter free and their field dependence is universal. At the end
of this section we discuss the role of time-reversal symmetry. In
Sec.\ \ref {Sec-SPhRelax} we obtain universal parameter-free
formulas for the rates of spin-phonon transitions between
tunnel-split states in the presence of the arbitrarily directed
magnetic field. Cases of the tunnel splitting due to the crystal
field and due to transverse magnetic field is considered in Secs.
\ref{Sec-SPhRelax-anis} and \ref{Sec-SPhRelax-field},
respectively. Implications of our results for experiment are
discussed in Sec.\ \ref {Sec-Discus}. Some illustrations for
uniaxial and biaxial spin models with the magnetic field, as well
as in-depth study of the role of the time-reversal symmetry, are
presented in Appendices. In particular, in Appendix \ref{App-E}\
we show that the same expressions for the rates of relaxation
between tunnel-split states can be obtained by traditional methods
but the required effort greatly exceeds the effort of our new
method.

\section{Spin-lattice interaction}

\label{Sec-SpinLattice}

\subsection{Laboratory frame}

\label{Sec-HSPh}

In the absence of phonons, one can choose the coordinate system in
which in Eqs.\ (\ref{uniaxial})--(\ref{cubic}) $e_{\beta
}^{(\alpha )}=\delta _{\alpha
\beta },$ i.e., $\hat{H}_{A}=-DS_{z}^{2}$ etc. A transverse phonon, $\mathbf{%
u}(\mathbf{r})$, rotates the axes of the local crystal field, $(\mathbf{e}%
^{(1)},\mathbf{e}^{(2)},\mathbf{e}^{(3)})$. This rotation can be described
by
\begin{equation}
\delta \mathbf{\phi (\mathbf{r})=}\frac{1}{2}\nabla \times \mathbf{u}(%
\mathbf{r})  \label{rotation}
\end{equation}
and it is performed by the $(3\times 3)$ rotation matrix ${\Bbb{R}}$,
\begin{equation}
\mathbf{e}^{(\alpha )}\rightarrow {\Bbb{R}}\mathbf{e}^{(\alpha )}\,,
\label{rotation-e}
\end{equation}
i.e., $e_{\beta }^{(\alpha )}\rightarrow {\Bbb{R}}_{\beta \beta ^{\prime
}}e_{\beta ^{\prime }}^{(\alpha )}.$ For small $\delta \mathbf{\phi }$, one
has
\begin{equation}
{\Bbb{R}}_{\alpha \beta }=\delta _{\alpha \beta }-\epsilon _{\alpha \beta
\gamma }\delta \phi _{\gamma }.  \label{rotation-matrix}
\end{equation}
We now notice that due to the rotational invariance of $\hat{H}_{A}$, the
rotation of the local frame $(\mathbf{e}^{(1)},\mathbf{e}^{(2)},\mathbf{e}%
^{(3)})$ is equivalent to the rotation of the vector $\mathbf{S}$ in the
opposite direction, $\mathbf{S}\rightarrow {\Bbb{R}}^{-1}\mathbf{S}$. As is
known, \cite{mes76} this rotation can be equivalently performed by the $%
(2S+1)\times (2S+1)$ matrix in the spin space,
\begin{equation}
\mathbf{S}\rightarrow \hat{R}\mathbf{S}\hat{R}^{-1}\,,\qquad \hat{R}=e^{-i%
\mathbf{S}\cdot \delta \mathbf{\phi }}\,.  \label{S-rotation}
\end{equation}
The total Hamiltonian can be written in the form
\begin{equation}
\hat{H}=\hat{R}\hat{H}_{A}\hat{R}^{-1}\mathbf{+}\hat{H}_{Z}+\hat{H}_{\mathrm{%
ph}},  \label{Hfull}
\end{equation}
where $\hat{H}_{A}$ is the crystal-field Hamiltonian in the absence of
phonons,
\begin{equation}
\hat{H}_{Z}=-g\mu _{B}\mathbf{H\cdot S}  \label{HZDef}
\end{equation}
is the Zeeman Hamiltonian, and $\hat{H}_{\mathrm{ph}}$ is the Hamiltonian of
harmonic phonons.

In the above formulas, $\mathbf{u}$ and $\mathbf{\phi }$ must be
treated as operators. Canonical quantization of phonons and Eq.\
(\ref{rotation}) give
\begin{eqnarray}
\mathbf{u} &=&\sqrt{\frac{\hbar }{2MN}}\sum_{\mathbf{k}\lambda }\frac{%
\mathbf{e}_{\mathbf{k}\lambda }e^{i\mathbf{k\cdot r}}}{\sqrt{\omega _{%
\mathbf{k}\lambda }}}\left( a_{\mathbf{k}\lambda }+a_{-\mathbf{k}\lambda
}^{\dagger }\right)  \label{uQuantized} \\
\delta \mathbf{\phi } &=&\frac{1}{2}\sqrt{\frac{\hbar }{2MN}}\sum_{\mathbf{k}%
\lambda }\frac{\left[ i\mathbf{k}\times \mathbf{e}_{\mathbf{k}\lambda }%
\right] e^{i\mathbf{k\cdot r}}}{\sqrt{\omega _{\mathbf{k}\lambda }}}\left(
a_{\mathbf{k}\lambda }+a_{-\mathbf{k}\lambda }^{\dagger }\right) ,
\label{deltaphiPh}
\end{eqnarray}
where $M$ is the mass of the unit cell, $N$ is the number of cells in the
crystal, $\mathbf{e}_{\mathbf{k}\lambda }$ are unit polarization vectors, $%
\lambda =t_{1},t_{2},l$ denotes polarization, and $\omega
_{k\lambda }=v_{\lambda }k$ is the phonon frequency. In
application to rigid magnetic clusters, Eqs.\ (\ref{uQuantized})
and (\ref{deltaphiPh}) describe quantized long wave phonons in the
elastic environment of the cluster. In the linear order in phonon
amplitudes one obtains
\begin{equation}
\hat{R}\hat{H}_{A}\hat{R}^{-1}\mathbf{\cong }\hat{H}_{A}\mathbf{+}\hat{H}_{%
\mathrm{s-ph}},\qquad \hat{H}_{\mathrm{s-ph}}=i\left[ \hat{H}_{A},\mathbf{S}%
\right] \cdot \delta \mathbf{\phi .}  \label{Smallphi}
\end{equation}
The total Hamiltonian can be written as
\begin{equation}
\hat{H}=\hat{H}_{0}+\hat{H}_{\mathrm{s-ph}},  \label{Hfullseparated}
\end{equation}
where $\hat{H}_{0}$ is the Hamiltonian of non-interacting spin and phonons,
\begin{equation}
\hat{H}_{0}=\hat{H}_{S}+\hat{H}_{\mathrm{ph}},  \label{H-0}
\end{equation}
and
\begin{equation}  \label{H-S}
\hat{H}_{S}=\hat{H}_{A}+\hat{H}_{Z},  \label{HSDef}
\end{equation}
is the spin Hamiltonian.

\subsection{Lattice frame}

\label{Sec-HSPhLatt}

The anisotropy Hamiltonian $\hat{H}_{A}$ is defined in the lattice
frame and thus, in this frame, it is not changed by the lattice
rotations. Unitary transformation to the lattice frame corresponds
to the rotation of $\hat{H}$ of Eq.\ (\ref{Hfull}) by the angle
$\delta \mathbf{\phi }$:
\begin{equation}
\hat{H}^{\prime }=\hat{R}^{-1}\hat{H}\hat{R}=\hat{H}_{A}+\hat{H}_{Z}^{\prime
}+\hat{H}_{\mathrm{ph}}^{\prime }.  \label{HLatt}
\end{equation}
Here
\begin{equation}
\hat{H}_{\mathrm{ph}}^{\prime }=\hat{R}^{-1}\hat{H}_{\mathrm{ph}}\hat{R}%
\cong \hat{H}_{\mathrm{ph}}-i\left[ \hat{H}_{\mathrm{ph}},\delta \mathbf{%
\phi }\right] \mathbf{\cdot S,}  \label{HphTrans}
\end{equation}
and
\begin{eqnarray}
\hat{H}_{Z}^{\prime } &=&\hat{R}^{-1}\hat{H}_{Z}\hat{R}\cong \hat{H}_{Z}-i%
\left[ \hat{H}_{Z},\mathbf{S}\right] \cdot \delta \mathbf{\phi }  \nonumber
\\
&=&\hat{H}_{Z}-g\mu _{B}\left[ \mathbf{H\times }\delta \mathbf{\phi }\right]
\mathbf{\cdot S,}  \label{HZTrans}
\end{eqnarray}
where we have used $\left[ \left( \mathbf{A\cdot S}\right) ,\left( \mathbf{%
B\cdot S}\right) \right] =i\mathbf{S\cdot }\left[ \mathbf{A\times B}\right]
. $

The full Hamiltonian in the lattice frame up to the first order in $\delta
\mathbf{\phi }$ is thus
\begin{equation}  \label{full-lattice}
\hat{H}^{\prime }=\hat{H}_{0}+\hat{H}_{\mathrm{s-ph}}^{\prime },
\label{HeffRearr}
\end{equation}
where $\hat{H}_{0}$ is given by Eq.\ (\ref{H-0}) and
\begin{equation}
\hat{H}_{\mathrm{s-ph}}^{\prime }=-i\left[ \hat{H}_{\mathrm{ph}},\delta
\mathbf{\phi }\right] \mathbf{\cdot S}-g\mu _{B}\left[ \mathbf{H\times }%
\delta \mathbf{\phi }\right] \mathbf{\cdot S.}  \label{s-ph'}
\end{equation}
With account of the relation
\begin{equation}
\,\delta \mathbf{\dot{\phi}}=\frac{i}{{\hbar }}[\hat{H}_{\mathrm{ph}},\delta
\mathbf{\phi }]  \label{phi-dot}
\end{equation}
one obtains
\begin{equation}
\hat{H}_{\mathrm{s-ph}}^{\prime }=-\hbar \hat{\mathbf{\Omega }}\cdot \mathbf{%
S},  \label{Hrot}
\end{equation}
where
\begin{equation}
\hat{\mathbf{\Omega }}=\delta \dot{\mathbf{\phi }}+\gamma \left[ \mathbf{H}%
\times \delta \mathbf{\phi }\right] ,  \label{OmegaDef}
\end{equation}
and $\gamma =g\mu _{B}/{\hbar }$ is the gyromagnetic ratio for $\mathbf{S}$.

Note that in the absence of the magnetic field the spin-lattice interaction
in the lattice frame can be obtained by simply writing \cite
{chu94prl,chumar02prb,chu04prl,chugar04prl} $\hat{H}^{\prime }=\hat{H}%
_{0}-\hbar \delta \dot{\mathbf{\phi }}\cdot \mathbf{S}$. The term $-\hbar
\delta \dot{\mathbf{\phi }}\cdot \mathbf{S}$ is of kinematic origin: in the
rotating coordinate frame the rotation is equivalent to the magnetic field $%
\mathbf{H}_{\mathrm{eff}}\mathbf{=}\delta
\mathbf{\dot{\phi}}/\gamma $ acting on the spin. The second term
in Eq.\ (\ref{OmegaDef}) describes the fact that the external
magnetic field, which is constant in the laboratory frame, makes
rotation in the lattice frame due to the transverse phonon.

\section{Matrix elements of spin-lattice interaction}

\label{Sec-ME}

We study spin-phonon transitions between the eigenstates of $\hat{H}_{0}$
that are direct products of the spin and phonon states,
\begin{equation}
\left| \Psi _{\pm }\right\rangle =\left| \psi _{\pm }\right\rangle \otimes
\left| \phi _{\pm }\right\rangle .  \label{psi-phi}
\end{equation}
Here $\left| \psi _{\pm }\right\rangle $ are the eigenstates of $\hat{H}_{S}$
with eigenvalues $E_{\pm }$ ($E_{+}>E_{-}$) and $\left| \phi _{\pm
}\right\rangle $ are the eigenstates of $\hat{H}_{\mathrm{ph}}$ with
energies $E_{\mathrm{ph},\pm }$. For $\hat{H}_{\mathrm{s-ph}}$ linear in
phonon amplitudes, the states $\left| \phi _{\pm }\right\rangle $ differ by
one emitted or absorbed phonon with a wave vector $\mathbf{k}$. Thus we will
use the designations
\begin{equation}
\left| \phi _{+}\right\rangle \equiv \left| n_{\mathbf{k}}\right\rangle
,\qquad \left| \phi _{-}\right\rangle \equiv \left| n_{\mathbf{k}%
}+1\right\rangle .  \label{phiviank}
\end{equation}
Spin-phonon transitions conserve energy:
\begin{equation}
E_{+}+E_{\mathrm{ph},+}=E_{-}+E_{\mathrm{ph},-}.  \label{Energyconservation}
\end{equation}
We calculate spin-phonon relaxation rates in both lattice frame and
laboratory frame and show that the result is the same, as physically
expected.

\subsection{Lattice frame}

\label{Sec-ME-Latt}

We first calculate the matrix element corresponding to the decay
of the spin $\left| \psi _{+}\right\rangle \rightarrow $ $\left|
\psi _{-}\right\rangle $ in the lattice frame. From Eq.\
(\ref{Hrot}) one obtains
\begin{equation}
\left\langle \Psi _{-}\left| \hat{H}_{\mathrm{s-ph}}^{\prime }\right| \Psi
_{+}\right\rangle =-\hbar {\mathbf{\Omega }}_{-+}\cdot \left\langle \psi
_{-}\left| \mathbf{S}\right| \psi _{+}\right\rangle ,  \label{MatrElLatRes}
\end{equation}
where
\begin{equation}
{\mathbf{\Omega }}_{-+}\equiv \left\langle \phi _{-}\left| \,\hat{\mathbf{%
\Omega }}\right| \phi _{+}\right\rangle .  \label{OmegatilDef}
\end{equation}
To calculate the matrix element ${\mathbf{\Omega }}_{-+},$ it is
convenient to step back and use the commutator form of $\delta
\dot{\mathbf{\phi }}$ given by Eq.\ (\ref{phi-dot}). One has
\begin{equation}
\left\langle \phi _{-}\left| \left[ \hat{H}_{\mathrm{ph}},\delta \mathbf{%
\phi }\right] \right| \phi _{+}\right\rangle =\left( E_{\mathrm{ph},-}-E_{%
\mathrm{ph},+}\right) \delta \mathbf{\phi }_{-+},  \label{PhononcommME}
\end{equation}
where
\begin{equation}
\delta \mathbf{\phi }_{-+}\equiv \left\langle \phi _{-}\left| \delta \mathbf{%
\phi }\right| \phi _{+}\right\rangle \,.  \label{phi12}
\end{equation}
It follows from the energy conservation, Eq.\
(\ref{Energyconservation}), that
\begin{equation}
E_{\mathrm{ph},-}-E_{\mathrm{ph},+}=E_{+}-E_{-}\equiv \hbar \omega _{0}.
\label{energyconservation}
\end{equation}
Thus one finally obtains
\begin{equation}
{\mathbf{\Omega }}_{-+}=i\omega _{0}\delta \mathbf{\phi }_{-+}+\gamma \left[
\mathbf{H}\times \delta \mathbf{\phi }_{-+}\right] .
\label{OmegaplusminusRes}
\end{equation}

\subsection{Laboratory frame}

\label{Sec-ME-Lab}

To check the consistency of our method, let us now obtain the
expression for the matrix element in the laboratory frame. Eq.\
(\ref{Smallphi}) gives
\begin{equation}
\left\langle \Psi _{-}\left| \hat{H}_{\mathrm{s-ph}}\right| \Psi
_{+}\right\rangle =i\left\langle \psi _{-}\left| \left[ \hat{H}_{A},\mathbf{S%
}\right] \right| \psi _{+}\right\rangle \cdot \delta \mathbf{\phi }_{-+}.
\label{MatrElLabDef}
\end{equation}
It is convenient to avoid explicitly working out the commutator in the spin
matrix element. To this end, we add and subtract $\hat{H}_{Z}$:
\begin{eqnarray}
&&\left\langle \psi _{-}\left| \left[ \hat{H}_{A},\mathbf{S}\right] \right|
\psi _{+}\right\rangle =\left\langle \psi _{-}\left| \left[ \hat{H}_{S},%
\mathbf{S}\right] \right| \psi _{+}\right\rangle  \nonumber \\
&&\qquad \qquad {}-i\left\langle \psi _{-}\left| \left[ \hat{H}_{Z},\mathbf{S%
}\right] \right| \psi _{+}\right\rangle .  \label{MatrElLabDefSubtr}
\end{eqnarray}
Now we can take into account that the states $\left| \psi _{\pm
}\right\rangle $ are exact eigenstates of the spin Hamiltonian
$\hat{H}_{S}$ of Eq.\ (\ref{HSDef}) with energies $E_{\pm }$:
\begin{equation}
\left\langle \psi _{-}\left| \left[ \hat{H}_{S},\mathbf{S}\right] \right|
\psi _{+}\right\rangle =\left( E_{-}-E_{+}\right) \left\langle \psi
_{-}\left| \mathbf{S}\right| \psi _{+}\right\rangle .
\label{CommDisappeares}
\end{equation}
The Zeeman term in Eq.\ (\ref{MatrElLabDefSubtr}) can be done as
in Eq.\ (\ref {HZTrans}). With the help of Eq.\
(\ref{energyconservation}) one then obtains
\begin{equation}
\left\langle \Psi _{-}\left| \hat{H}_{\mathrm{s-ph}}\right| \Psi
_{+}\right\rangle =-\hbar {\mathbf{\Omega }}_{-+}\cdot \left\langle \psi
_{-}\left| \mathbf{S}\right| \psi _{+}\right\rangle ,  \label{MatrElLabRes}
\end{equation}
where ${\mathbf{\Omega }}_{-+}$ is given by Eq.\
(\ref{OmegaplusminusRes}). We see that the spin-phonon matrix
elements computed in the laboratory and lattice frames, Eqs.\
(\ref{MatrElLatRes}) and (\ref{MatrElLabRes}), are exactly the
same.

\section{Spin matrix elements for tunnel-split states}

\label{Sec-MatrElem}

\begin{figure}[t]
\unitlength1cm
\begin{picture}(11,9)
\centerline{\psfig{file=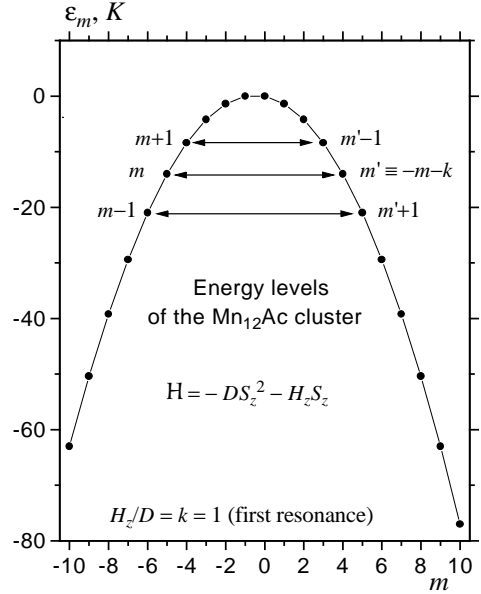,angle=0,width=8cm}}
\end{picture}
\caption{Spin energy levels of a Mn$_{12}$Ac molecule for $H_{x}=0$ and $%
H_{z}=D$ corresponding to the first resonance, $k=1$. }
\label{fig_levels}
\end{figure}

The method of the computation of transition rates outlined in the previous
sections has significant advantage over conventional methods. Regardless of
the explicit form of the spin Hamiltonian, only the matrix elements of the
operator $\mathbf{S}$ need to be computed. Especially interesting is the
case of strong uniaxial anisotropy, in which $\hat{H}_{S}$ nearly commutes
with $S_{z},$ so that the energy levels can be approximately described with
the help of the quantum number $m$:
\begin{equation}
E_{m}=E_{m}^{(A)}-g\mu _{B}H_{z}m,  \label{EmDef}
\end{equation}
where $E_{m}^{(A)}$ is the contribution of the crystal field that satisfies $%
E_{-m}^{(A)}=E_{m}^{(A)}$. The structure of the energy levels for
this model is shown in Fig.\ \ref{fig_levels}.
The two levels $m$ and $m^{\prime }$ are in resonance for the values of the
magnetic field
\begin{equation}
g\mu _{B}H_{z,mm^{\prime }}^{(\mathrm{res})}=\frac{E_{m}^{(A)}-E_{m^{\prime
}}^{(A)}}{m-m^{\prime }}\equiv \frac{\hbar \omega _{mm^{\prime }}^{(A)}}{%
m-m^{\prime }}.  \label{HzResDef}
\end{equation}
For Mn$_{12}$ the dominant term in the uniaxial anisotropy energy is $%
E_{m}^{(A)}=-Dm^{2}$ and thus $g\mu _{B}H_{z,mm^{\prime }}^{(\mathrm{res}%
)}=-D(m+m^{\prime })=Dk,$ where $k$ is the resonance number (see
Fig.\ \ref {fig_levels}). The level bias is given by
\begin{eqnarray}
W &\equiv &E_{m}-E_{m^{\prime }}=g\mu _{B}\left( H_{z}-H_{z,mm^{\prime }}^{(%
\mathrm{res})}\right) \left( m^{\prime }-m\right)  \nonumber \\
&=&\left( m^{\prime }-m\right) g\mu _{B}H_{z}+\hbar \omega _{mm^{\prime
}}^{(A)}.  \label{WDef}
\end{eqnarray}
Choosing $dW/dH_{z}>0$ for certainty, makes $m^{\prime }>m$ in all
subsequent calculations.

\subsection{The two-state model: $\left\langle \protect\psi _{-}\left|
S_{z}\right| \protect\psi _{+}\right\rangle $}

\label{Sec-MatrElem-twostate}

Due to the terms in $\hat{H}_{S}$ that do not commute with $S_{z},$ the true
eigenstates of $\hat{H}_{S}$ are expansions over the complete $\left|
m\right\rangle $ basis:
\begin{equation}
\left| \psi _{\pm }\right\rangle =\sum_{m^{\prime \prime }=-S}^{S}c_{\pm
,m}\left| m^{\prime \prime }\right\rangle .  \label{PsikGen}
\end{equation}
If one neglects tunneling (i. e., hybridization of the states $\left|
m\right\rangle $ and $\left| m^{\prime }\right\rangle $ on different sides
of the potential barrier, see Fig.\ \ref{fig_levels}), each pure state $%
\left| m\right\rangle $ should be replaced by $\left| \tilde{m}\right\rangle
$:
\begin{equation}
\left| \tilde{m}\right\rangle =\sum_{m^{\prime \prime }=-S}^{S}c_{mm^{\prime
\prime }}\left| m^{\prime \prime }\right\rangle ,  \label{mmprimetildeDef}
\end{equation}
where $\left| c_{mm}\right| \cong 1$ and all other coefficients are small.
Hybridization of the states $\left| \tilde{m}\right\rangle $ and $\left|
\tilde{m}^{\prime }\right\rangle $ can be taken into account in the
framework of the two-state model
\begin{eqnarray}
\left\langle \tilde{m}_{i}\left| \hat{H}_{S}\right| \tilde{m}%
_{i}\right\rangle &=&E_{m_{i}},\qquad m_{i}=m,m^{\prime }  \nonumber \\
\left\langle \tilde{m}\left| \hat{H}_{S}\right| \tilde{m}^{\prime
}\right\rangle &=&\frac{1}{2}\Delta e^{i\varphi },  \label{TLS}
\end{eqnarray}
where $\Delta $ is the tunnel splitting of the levels $m$ and $m^{\prime }$
that can be calculated from the exact spin Hamiltonian $\hat{H}_{S}$ or
determined experimentally and $\varphi $ is a phase. If $\hat{H}_{S}$
expressed in terms of $S_{z}$ and $S_{\pm }$ is real, then the only two
possibilities are $\varphi =0,\pi ,$ so that $e^{i\varphi }=\pm 1.$
Diagonalizing this $2\times 2$ matrix yields the eigenvalues
\begin{equation}
E_{\pm }=\frac{1}{2}\left( E_{m}+E_{m^{\prime }}\pm \sqrt{W^{2}+\Delta ^{2}}%
\right) .  \label{Epm}
\end{equation}
The energy difference is
\begin{equation}
E_{+}-E_{-}\equiv \hbar \omega _{0}\equiv \sqrt{W^{2}+\Delta ^{2}}.
\label{hbaromegaDef}
\end{equation}
The corresponding eigenvectors can be represented in the form
\begin{equation}
\left| \psi _{\pm }\right\rangle =\frac{1}{\sqrt{2}}\left( C_{\pm
}e^{i\varphi /2}\left| \tilde{m}\right\rangle \pm C_{\mp }e^{-i\varphi
/2}\left| \tilde{m}^{\prime }\right\rangle \right) ,  \label{Psipm}
\end{equation}
where
\begin{equation}
C_{\pm }=\sqrt{1\pm \frac{W}{\sqrt{W^{2}+\Delta ^{2}}}}.  \label{Cpm}
\end{equation}
Far from the resonance, $\left| W\right| \gg \Delta ,$ the eigenstates and
energy eigenvalues reduce to those of $\left| \tilde{m}\right\rangle $ \ and
$\left| \tilde{m}^{\prime }\right\rangle $ states. For $W>0$ and $W\gg
\Delta $ one has $\left| \psi _{+}\right\rangle \cong e^{i\varphi /2}\left|
\tilde{m}\right\rangle $ and $\left| \psi _{-}\right\rangle \cong
-e^{-i\varphi /2}\left| \tilde{m}^{\prime }\right\rangle .$ On the contrary,
for $W<0$ and $\left| W\right| \gg \Delta $ one has $\left| \psi
_{+}\right\rangle \cong e^{-i\varphi /2}\left| \tilde{m}^{\prime
}\right\rangle $ and $\left| \psi _{-}\right\rangle \cong e^{i\varphi
/2}\left| \tilde{m}\right\rangle .$ Exactly at the resonance, $W=0,$ the
eigenstates $\left| \psi _{\pm }\right\rangle $ are superpositions of $%
\left| \tilde{m}\right\rangle $ \ and $\left| \tilde{m}^{\prime
}\right\rangle $ with equal weights, and signs depending on the phase $%
\varphi .$

Eq.\ (\ref{Psipm}) is sufficient to calculate the matrix element
\begin{equation}
\left\langle \psi _{-}\left| S_{z}\right| \psi _{+}\right\rangle =-\frac{%
\Delta }{\sqrt{W^{2}+\Delta ^{2}}}\frac{m^{\prime }-m}{2}.  \label{MatrElSz}
\end{equation}
This result does not depend on the detailed form of $\hat{H}_{A},$
provided that the condition of strong uniaxial anisotropy is
fulfulled, and it is universal in this sense. It is large near the
resonance, $W^{2}\lesssim \Delta ^{2}$, but it becomes small far
from the resonance. This the only matrix element that plays a role
in the relaxation for $H=0,$ as will be shown in Sec.\
\ref{Sec-MatrElem-TR}. Also for $H\neq 0$ in the case where the
tunnel splitting $\Delta $ is due to the transverse anisotropy,
this
matrix element is dominant. In another particular case when $\left[ \hat{H}%
_{A},S_{z}\right] =0$ and thus the only source of the splitting $\Delta $ is
the transverse field one has to take into account matrix elements $%
\left\langle \psi _{-}\left| S_{\pm }\right| \psi _{+}\right\rangle .$ This
will be done in the next subsection.

\subsection{Beyond the two-state model: $\left\langle \protect\psi
_{-}\left| S_{\pm }\right| \protect\psi _{+}\right\rangle $}

\label{Sec-MatrElem-beyond}

Consider the spin Hamiltonian $\hat{H}_{S}$ of Eq.\ (\ref{HSDef}), where $%
\hat{H}_{A}$ satisfies $\left[ \hat{H}_{A},S_{z}\right] =0$ and is strong in
comparison to $\hat{H}_{Z}.$ Due to this rotational invariance the
orientation of the transverse field is unimportant, so we direct it along
the $x$ axis for simplicity:
\begin{equation}
\hat{H}_{Z}=-g\mu _{B}H_{z}S_{z}-g\mu _{B}H_{x}S_{x}  \label{HamHxHz}
\end{equation}
and asssume $H_{x}>0.$ For strong uniaxial anisotropy $\hat{H}_{A}$ one can
obtain the splitting of resonance levels $\left| m\right\rangle $ and $%
\left| m^{\prime }\right\rangle $ ($m^{\prime }>m$) perturbatively in $H_{x}$%
:
\begin{equation}
\frac{\Delta }{2}=\left| V_{m,m+1}\frac{1}{E_{m+1}-E_{m}}V_{m+1,m+2}\ldots
V_{m^{\prime }-1,m^{\prime }}\right| ,  \label{DeltaHChain}
\end{equation}
where
\begin{equation}
V_{m,m+1}=-\frac{1}{2}g\mu _{B}H_{x}\left\langle m\left| S_{-}\right|
m+1\right\rangle .  \label{VChain}
\end{equation}
In the particular case $\hat{H}_{A}=-DS_{z}^{2}$ the calculation
in Eq.\ (\ref {DeltaHChain}) yields \cite{gar91jpa}
\begin{eqnarray}
\Delta &=&\frac{2D}{\left[ \left( m^{\prime }-m-1\right) !\right] ^{2}}
\nonumber \\
&&\times \sqrt{\frac{\left( S+m^{\prime }\right) !\left( S-m\right) !}{%
\left( S-m^{\prime }\right) !\left( S+m\right) !}}\left( \frac{g\mu _{B}H_{x}%
}{2D}\right) ^{m^{\prime }-m}.  \label{DeltaPert}
\end{eqnarray}
One can see from this calculation that\ in our case\ in Eq.\
(\ref{TLS}) is \
\begin{equation}
e^{i\varphi }=\left( -1\right) ^{m^{\prime }-m}  \label{eiphiHx}
\end{equation}
for $H_{x}>0.$

To compute matrix elements of $S_{\pm }$ between the tunnel-split states $%
\left| \psi _{\pm }\right\rangle $ one has to go beyond the two-state model
of the preceding section. Taking into account that the states $\left| \tilde{%
m}\right\rangle $ \ and $\left| \tilde{m}^{\prime }\right\rangle $ are not
pure $\left| m\right\rangle $ \ and $\left| m^{\prime }\right\rangle $
states, see Eq.\ (\ref{mmprimetildeDef}), one obtains small values for $%
\left\langle \tilde{m}\left| S_{-}\right| \tilde{m}^{\prime }\right\rangle $
and $\left\langle \tilde{m}^{\prime }\left| S_{+}\right| \tilde{m}%
\right\rangle $ that are, however, essential in the spin-phonon relaxation
in the case when $\Delta $ is caused solely by the transverse field. These
small terms can be calculated perturbatively by building chains of
elementary matrix elements that join the pure states, $\left| m\right\rangle
$ and $\left| m^{\prime }\right\rangle $. These chains contain terms of $%
\hat{H}_{S}$ that do not commute with $S_{z}$ and the
corresponding energy denominators, similar to the perturbative
chain of Eq.\ (\ref{DeltaHChain}). The difference is that the
elementary matrix element with the ``external'' operator $S_{-}$
does not contain $-(1/2)g\mu _{B}H_{x},$ unlike all other
elementary matrix elements, and the external operator $S_{-}$ can
be inserted into the chain at $m^{\prime }-m$ different positions.
For so defined indirect (real!) matrix elements one obtains
\begin{equation}
\left\langle \tilde{m}\left| S_{-}\right| \tilde{m}^{\prime }\right\rangle
=\left\langle \tilde{m}^{\prime }\left| S_{+}\right| \tilde{m}\right\rangle
=\left( -1\right) ^{m^{\prime }-m+1}\frac{\left( m^{\prime }-m\right) \Delta
}{g\mu _{B}H_{x}}.  \label{MSpmS}
\end{equation}
Note that this result does not use the specific form of $\hat{H}_{A}$ and it
is thus universal, similarly to Eq.\ (\ref{MatrElSz}). For $S=1/2$ one has $%
\Delta =g\mu _{B}H_{x}$ and $\left\langle -S\left| S_{-}\right|
S\right\rangle =1$, which is a correct result. Calculating the
matrix elements between the tunnel-split states $\left| \psi _{\pm
}\right\rangle $ given by Eq.\ (\ref{Psipm}) with $m\Rightarrow
\tilde{m},$ $m^{\prime }\Rightarrow \tilde{m}^{\prime },$ and
$\varphi =0$ one obtains the unchanged result for $\left\langle
\psi _{-}\left| S_{z}\right| \psi _{+}\right\rangle $ that is
given by Eq.\ (\ref{MatrElSz}). For transverse operators one
obtains
\begin{equation}
\left\langle \psi _{-}\left| S_{\pm }\right| \psi _{+}\right\rangle =\frac{%
\left( m^{\prime }-m\right) \Delta }{2g\mu _{B}H_{x}}\left( \pm 1+\frac{W}{%
\sqrt{W^{2}+\Delta ^{2}}}\right) .  \label{MSpm}
\end{equation}
Below we will need
\begin{equation}
\left\langle \psi _{-}\left| S_{x}\right| \psi _{+}\right\rangle =\frac{%
\left( m^{\prime }-m\right) \Delta }{2g\mu _{B}H_{x}}\frac{W}{\sqrt{%
W^{2}+\Delta ^{2}}}  \label{MSx}
\end{equation}
and
\begin{equation}
\left\langle \psi _{-}\left| S_{y}\right| \psi _{+}\right\rangle =-i\frac{%
\left( m^{\prime }-m\right) \Delta }{2g\mu _{B}H_{x}}.  \label{MSy}
\end{equation}
Note that these results are only valid if $\Delta $ is due to the transverse
field. When $\left[ \hat{H}_{A},S_{z}\right] \neq 0$ the dominating source
of $\Delta $ is $\hat{H}_{A}$. In this case the matrix elements $%
\left\langle \psi _{-}\left| S_{\pm }\right| \psi _{+}\right\rangle $ are
much smaller than $\left\langle \psi _{-}\left| S_{z}\right| \psi
_{+}\right\rangle $ and they can be safely neglected.

\subsection{Role of time-reversal symmetry}

\label{Sec-MatrElem-TR}

Let us now discuss the role of time reversal symmetry for spin-phonon
transitions. For the total Hamiltonian H to be invariant under time
reversal, the external field $\mathbf{H}$ must be zero. Thus we will
consider only this case here. For $\mathbf{H}=0,$ tunneling can only arise
from the transverse anisotropy which lifts the degeneracy of the eigenstates
$|m\rangle $ and $|-m\rangle $ of the longitudinal part of the crystal field
Hamiltonian $\hat{H}_{A}$. According to the Kramers' theorem, this
degeneragy is lifted only for integer spins $S.$ Let $|\psi _{\pm }\rangle $
denote the corresponding tunnel-split eigenstates of $\hat{H}_{A}$. It will
be shown in Appendix \ref{App-TR} that $|\psi _{\pm }\rangle $ are
eigenstates of the time reversal operator $\hat{K}$ with eigenvalues $\pm 1$%
, i.e. $|\psi _{+}\rangle $ and $|\psi _{-}\rangle $ have \emph{opposite}
parity with respect to time reversal. The spin-lattice Hamiltonian $\hat{H}_{%
\mathrm{s-ph}}$ in the laboratory frame that is given by Eq.\ (\ref{Smallphi}%
) is invariant under time reversal. It will be proven in Appendix \ref
{App-TR} that this property, together with the \emph{antiunitary} character
of $\hat{K},$ leads to the relation

\begin{equation}
\langle \psi _{-}|\hat{H}_{\mathrm{s-ph}}|\psi _{+}\rangle =-\langle \psi
_{-}|\hat{H}_{\mathrm{s-ph}}|\psi _{+}\rangle ^{\ast }  \label{eq1}
\end{equation}
for the spin matrix element in the case of integer $S$. One can
see that spin-phonon transitions are not ruled out by the
time-reversal symmetry if the matrix element is imaginary. Complex
conjugation in Eq.\ (\ref{eq1}) makes the situation different from
the cases of spatial symmetries.

A specific example showing the absence of a time-reversal
selection rule is the crystal field Hamiltonian $\hat{H}_{A}$ from
Eq.\ (\ref{biaxial}). The contribution to
$\hat{H}_{\mathrm{s-ph}}$ from the phonons rotating the lattice
around $z$-axis is imaginary and is given by Eq.\ (\ref{HsphE}).
On
the other hand, eigenfunctions $\left| \psi _{\pm }\right\rangle $ of Eq.\ (%
\ref{Psipm}) are real for $E>0$ [see comment below Eq.\
(\ref{DeltaERes})]. Thus $\left\langle \psi _{-}\left|
\hat{H}_{\mathrm{s-ph}}\right| \psi
_{+}\right\rangle $ is imaginary and Eq.\ (\ref{eq1}) is satisfied by $%
\left\langle \psi _{-}\left| \hat{H}_{\mathrm{s-ph}}\right| \psi
_{+}\right\rangle \neq 0$.\newline

In the lattice frame, as well as in the laboratory frame, if our new method
is used, one has to calculate matrix elements of the spin operator $\mathbf{S%
}$ [see Eqs.\ (\ref{MatrElLatRes}) and (\ref{MatrElLabRes})]
between the states $|\psi _{+}\rangle $ and $|\psi _{-}\rangle $.
In contrast to the spin part of $\hat{H}_{\mathrm{s-ph}},$ the
operator $\mathbf{S}$ breaks time-reversal symmetry, see Eq.\
(\ref{eq3}). Thus one obtains the relation
\begin{equation}
\langle \psi _{-}|\mathbf{S}|\psi _{+}\rangle =\langle \psi _{-}|\mathbf{S}%
|\psi _{+}\rangle ^{\ast }  \label{STRev}
\end{equation}
instead of Eq.\ (\ref{eq1}). For the biaxial model with $E>0$ the states $%
|\psi _{+}\rangle $ and $|\psi _{-}\rangle $ are real, and one obtains the
selection rule $\langle \psi _{-}|S_{y}|\psi _{+}\rangle =0$ because $%
S_{y}=(i/2)(S_{-}-S_{+})$ is imaginary. On the other hand, $S_{x}$ and $%
S_{z} $ are real, and time-reversal symmetry does not lead to selection
rules for them. However one obtains $\langle \psi _{-}|S_{x}|\psi
_{+}\rangle =0,$ too, that can be shown with the help of $%
S_{x}=(1/2)(S_{-}+S_{+})$ and Eq.\ (\ref{Psipm}). Thus the matrix element $%
\langle \psi _{-}|S_{z}|\psi _{+}\rangle $ given by Eq.\
(\ref{MatrElSz}), corresponding to phonons rotating the lattice
about the z-axis, is the only matrix element that is responsible
for the relaxation between the tunnel-split states of the spin
Hamiltonian for $H=0$.

\section{Spin-phonon relaxation for tunnel-split states}

\label{Sec-SPhRelax}

The rates of spin-phonon transitions can be calculated with the
help of the Fermi golden rule. One should distinguish between two
situations: when the tunnel splitting $\Delta $ is caused by the
terms in $\hat{H}_{A}$ that do not commute with $S_{z},$ such as
the transverse anisotropy in Eq.\ (\ref
{biaxial}), and when $\Delta $ is caused by the transverse field in $\hat{H}%
_{Z}$. The physical difference between these two cases is that $\hat{H}_{A}$
is defined in the lattice frame and thus it is rotated by the transverse
phonons, whereas $\hat{H}_{Z}$ is defined in the laboratory frame and it is
not rotated. We will see that in the first case, in the absence of the
field, relaxation is due to the phonons rotating the lattice around the $z$
axis. To the contrary, in the second case ($\hat{H}_{A}$ commutes with $%
S_{z} $) these phonons are decoupled from the spin and they produce no
effect.

\subsection{Tunneling induced by the anisotropy}

\label{Sec-SPhRelax-anis}

In this case, as was shown at the end of Sec.\
\ref{Sec-MatrElem-TR}, the transition matrix element in Eq.\
(\ref{MatrElLabRes}) is due to the operator $S_{z}$ only for
$H=0.$ For nonzero fields, $S_{z}$ provides the dominant
contribution to the matrix element because $\left\langle \psi
_{-}\left| S_{z}\right| \psi _{+}\right\rangle \sim 1$ in the
vicinity of the resonance [see Eq.\ (\ref{MatrElSz})], whereas
other matrix elements can be shown to be small. Thus one can write
\begin{equation}
\left\langle \Psi _{-}\left| \hat{H}_{\mathrm{s-ph}}\right| \Psi
_{+}\right\rangle =-\left\langle \psi _{-}\left| S_{z}\right| \psi
_{+}\right\rangle \hbar \Omega _{-+,z}\,.  \label{MatrElwPhonons}
\end{equation}
With the help of Eq.\ (\ref{deltaphiPh}) and Eq.\ (\ref{OmegatilDef}), Eq.\ (%
\ref{MatrElwPhonons}) can be rewritten as
\begin{eqnarray}
&&\left\langle \Psi _{-}\left| \hat{H}_{\mathrm{s-ph}}\right| \Psi
_{+}\right\rangle =  \nonumber \\
&&\frac{\hbar }{\sqrt{N}}\sum_{\mathbf{k}\lambda }V_{\mathbf{k}\lambda
}\left\langle n_{\mathbf{k}}+1\left| \left( a_{\mathbf{k}\lambda }+a_{-%
\mathbf{k}\lambda }^{\dagger }\right) \right| n_{\mathbf{k}}\right\rangle ,
\label{MatrElwPhononsquant}
\end{eqnarray}
where we used the designations of Eq.\ (\ref{phiviank}) and
\begin{eqnarray}
V_{\mathbf{k}\lambda } &\equiv &-\left\langle \psi _{-}\left| S_{z}\right|
\psi _{+}\right\rangle \frac{e^{i\mathbf{k\cdot r}}}{\sqrt{8M\hbar \omega _{%
\mathbf{k}\lambda }}}\left\{ \hbar \omega _{0}\left[ \mathbf{k}\times
\mathbf{e}_{\mathbf{k}\lambda }\right] _{z}\right.  \nonumber \\
&&\qquad \qquad +\left. ig\mu _{B}\left[ \mathbf{H}\times \left[ \mathbf{k}%
\times \mathbf{e}_{\mathbf{k}\lambda }\right] \right] _{z}\right\} .
\label{VklamDef}
\end{eqnarray}
The decay rate $W_{-+}$ of the upper spin state into the lower state,
accompanied by the emission of a phonon, and the rate $W_{+-}$ of the
inverse process are given by
\begin{equation}
\left\{
\begin{array}{c}
W_{-+} \\
W_{+-}
\end{array}
\right\} =W_{0}\left\{
\begin{array}{c}
n_{\omega _{0}}+1 \\
n_{\omega _{0}}
\end{array}
\right\} ,  \label{WpmRes}
\end{equation}
where $n_{\omega _{0}}=\left( e^{\beta \hbar \omega _{0}}-1\right) ^{-1}$
with $\beta =1/(k_{B}T)$ is the phonon occupation number at equilibrium and
\begin{equation}
W_{0}=\frac{1}{N}\sum_{\mathbf{k}\lambda }\left| V_{\mathbf{k}\lambda
}\right| ^{2}2\pi \delta \left( \omega _{\mathbf{k}\lambda }-\omega
_{0}\right) .  \label{WmpDef}
\end{equation}
The master equation for the populations of the spin states $n_{+}$ and $%
n_{-} $ satisfying $n_{+}+n_{-}=1$ and
\begin{eqnarray}
\dot{n}_{+} &=&-W_{-+}n_{+}+W_{+-}n_{-}  \nonumber \\
&=&-\Gamma n_{+}+W_{+-}  \label{MasterEq}
\end{eqnarray}
defines the relaxation rate
\begin{equation}
\Gamma =W_{+-}+W_{-+}=W_{0}\left( 2n_{\omega _{0}}+1\right) .
\label{GammaDef}
\end{equation}

One can see that the two terms Eq.\ (\ref{VklamDef}) do not
interfere and that only transverse phonons, $\lambda =t,$ are
active in the relaxation process. In the first term of Eq.\
(\ref{VklamDef}) one can use \ $\left[
\mathbf{k}\times \mathbf{e}_{\mathbf{k}t}\right] =\pm k\mathbf{e}_{\mathbf{k}%
t^{\prime }},$ where $t$ and $t^{\prime }$ denote different transverse
phonons. Summation over polarization vectors of transverse phonons and
averaging over the directions of $\mathbf{k}$ can be performed with the help
of the formulas given in the Appendix \ref{App-math}. Replacing $N^{-1}\sum_{%
\mathbf{k}}\ldots $ by $v_{0}\int d^{3}k/(2\pi )^{3}\ldots $
($v_{0}$ being the unit cell volume) and using Eq.\
(\ref{MatrElSz}) one obtains
\begin{eqnarray}
W_{0} &=&\frac{\left[ (m^{\prime }-m)/2\right] ^{2}}{12\pi \hbar }\frac{%
\Delta ^{2}}{Mv_{t}^{2}}\frac{\omega _{0}\left( \omega _{0}^{2}+\gamma
^{2}H_{\bot }^{2}\right) }{\omega _{D}^{3}}  \nonumber \\
&=&\frac{\left[ (m^{\prime }-m)/2\right] ^{2}}{12\pi \hbar }\frac{\Delta
^{2}\omega _{0}\left( \omega _{0}^{2}+\gamma ^{2}H_{\bot }^{2}\right) }{\rho
v_{t}^{5}},  \label{W0Res}
\end{eqnarray}
where $\rho $ is the mass density, ${\omega }_{D}\equiv v_{t}/v_{0}^{1/3}$
is the Debye frequency for the tranverse phonons, $\omega _{0}$ is given by
Eq.\ (\ref{hbaromegaDef}), and $H_{\bot }^{2}=H_{x}^{2}+H_{y}^{2}$. In Eq.\ (%
\ref{W0Res}) the term $\omega _{0}^{2}$ \ in the brackets is due to the
transverse phonons that rotate the lattice around the $z$ axis, whereas the
transverse-field term is due to the phonons that rotate the lattice around $%
x $ and $y$ axes.

The beauty of Eq.\ (\ref{W0Res}) is that it gives a universal
expression for the transition rate, which does not depend on the
exact form of the crystal-field Hamiltonian, provided that the
uniaxial anisotropy dominates, and is expressed entirely in terms
of independently measurable parameters. Equations (\ref{WDef}),
(\ref{hbaromegaDef}), and (\ref{W0Res}) show that for $\left(
m^{\prime }-m\right) ^{2}\gg 1$ the contribution of the
longitudinal bias field to the relaxation is much stronger than
the contribution of the transverse field.

\subsection{Tunneling induced by the transverse field}

\label{Sec-SPhRelax-field}

The spin matrix elements for this case have been computed in Sec.\
\ref {Sec-MatrElem-beyond}. The calculation of ${\mathbf{\Omega
}}_{-+}$ is more complicated than in the previous subsection
because one has to take into account the transverse components in
the field term in Eq.\ (\ref {OmegaplusminusRes}),
\begin{eqnarray}
{\mathbf{\Omega }}_{-+} &=&\left( i\omega _{0}\mathbf{e}_{x}+\gamma H_{z}%
\mathbf{e}_{y}\right) \delta \phi _{-+,x}  \nonumber \\
&&+\left( i\omega _{0}\mathbf{e}_{y}-\gamma H_{z}\mathbf{e}_{x}+\gamma H_{x}%
\mathbf{e}_{z}\right) \delta \phi _{-+,y}  \nonumber \\
&&+\left( i\omega _{0}\mathbf{e}_{z}-\gamma H_{x}\mathbf{e}_{y}\right)
\delta \phi _{-+,z}.  \label{Omegadphi}
\end{eqnarray}
The transition matrix element, Eqs.\ (\ref{MatrElLatRes}) or (\ref
{MatrElLabRes}), can be presented in the form
\begin{equation}
\left\langle \Psi _{-}\left| \hat{H}_{\mathrm{s-ph}}\right| \Psi
_{+}\right\rangle =K_{x}\delta \phi _{-+,x}+K_{y}\delta \phi
_{-+,y}+K_{z}\delta \phi _{-+,z}.  \label{Kdef}
\end{equation}
Using Eqs.\ (\ref{MatrElSz}) and (\ref{MSy}) one obtains
\begin{equation}
K_{z}=-i\left\langle \psi _{-}\left| S_{z}\right| \psi _{+}\right\rangle
\hbar \omega _{0}+\left\langle \psi _{-}\left| S_{y}\right| \psi
_{+}\right\rangle g\mu _{B}H_{x}=0.  \label{MatrEldphiz}
\end{equation}
This result was expected because phonons rotating the lattice around the $z$
axis cannot cause any effect for $\left[ \hat{H}_{A},S_{z}\right] =0.$ To
the contrary, $K_{x}$ and $K_{y}$ are nonzero:
\begin{eqnarray}
K_{x} &=&-i\left\langle \psi _{-}\left| S_{x}\right| \psi _{+}\right\rangle
\hbar \omega _{0}-g\mu _{B}H_{z}\left\langle \psi _{-}\left| S_{y}\right|
\psi _{+}\right\rangle  \nonumber \\
&=&-\frac{i\left( m^{\prime }-m\right) \Delta }{2g\mu _{B}H_{x}}\left(
W-g\mu _{B}H_{z}\right)  \label{Kxx}
\end{eqnarray}
and
\begin{eqnarray}
K_{y} &=&-i\left\langle \psi _{-}\left| S_{y}\right| \psi _{+}\right\rangle
\hbar \omega _{0}+g\mu _{B}H_{z}\left\langle \psi _{-}\left| S_{x}\right|
\psi _{+}\right\rangle  \nonumber \\
&&\qquad \qquad -g\mu _{B}H_{x}\left\langle \psi _{-}\left| S_{z}\right|
\psi _{+}\right\rangle  \nonumber \\
&=&-g\mu _{B}H_{x}S\frac{\left( m^{\prime }-m\right) \Delta }{2\hbar \omega
_{0}}  \nonumber \\
&&\qquad \times \left( -A+\frac{W\left( W-g\mu _{B}H_{z}\right) }{\left(
g\mu _{B}H_{x}\right) ^{2}}\right) ,  \label{Kyy}
\end{eqnarray}
where $A$ is given by
\begin{equation}
A\equiv 1-\frac{\Delta ^{2}}{\left( g\mu _{B}H_{x}\right) ^{2}}
\label{ABDef}
\end{equation}
and $W$ is related to $H_{z}$ by Eq.\ (\ref{WDef}).

Quantization of the phonon field, Eq.\ (\ref{deltaphiPh}), yields
Eq.\ (\ref {MatrElwPhononsquant}), where
\begin{equation}
V_{\mathbf{k}\lambda }=\frac{e^{i\mathbf{k\cdot r}}}{\sqrt{8M\hbar \omega _{%
\mathbf{k}\lambda }}}\left\{ K_{x}\left[ \mathbf{k}\times \mathbf{e}_{%
\mathbf{k}\lambda }\right] _{x}+K_{y}\left[ \mathbf{k}\times \mathbf{e}_{%
\mathbf{k}\lambda }\right] _{y}\right\} .  \label{Vkvv}
\end{equation}
The relaxation rates between the states $m$ and $m^{\prime }$ are
given by Eq.\ (\ref{WpmRes}), where
\begin{equation}
W_{0}=\frac{1}{12\pi \hbar }\frac{\left| K_{x}\right| ^{2}+\left|
K_{y}\right| ^{2}}{Mv_{t}^{2}}\frac{\omega _{0}^{3}}{\omega _{D}^{3}}
\label{W0KxKy}
\end{equation}
and $K_{x}$ and $K_{y}$ are given by Eqs.\ (\ref{Kxx}) and
(\ref{Kyy}), respectively. The general result is cumbersome but it
simplifies for the ground-state resonance, $m=-S$ and $m^{\prime
}=S$:
\begin{equation}
W_{0}=\frac{S^{2}}{12\pi \hbar }\frac{\Delta ^{2}}{Mv_{t}^{2}}\frac{\omega
_{0}\left( \gamma H_{\bot }\right) ^{2}}{\omega _{D}^{3}}Q,  \label{W0zQ}
\end{equation}
where we have replaced $H_{x}\Rightarrow H_{\bot },$ the field perpendicular
to the anisotropy axes, and
\begin{eqnarray}
Q &\equiv &\left[ 1-\left( \frac{\omega _{0}}{\gamma H_{\perp }}\right) ^{2}+%
\frac{1}{2S}\left( \frac{W}{g\mu _{B}H_{\perp }}\right) ^{2}\right] ^{2}
\nonumber \\
&+&\left( 1-\frac{1}{2S}\right) ^{2}\left( \frac{W}{g\mu _{B}H_{\perp }}%
\right) ^{2}\left( \frac{\omega _{0}}{\gamma H_{\perp }}\right) ^{2}.
\label{QDef}
\end{eqnarray}

In the case of $S=1/2$ one has $\Delta =g\mu _{B}H_{\bot },$ so that $Q$ and
$W_{0}$ are zero. This is expected as the crystal-field anisotropy $\hat{H}%
_{A}$ disappears for $S=1/2$ and phonons do not couple to the spin. In
general, for the ground-state splitting one has $\Delta \varpropto H_{\bot
}^{2S}$ and $\Delta \ll g\mu _{B}H_{\bot }$ for any $S>1/2.$

\section{Discussion}

\label{Sec-Discus}

We have studied a universal mechanism of the relaxation of a spin
state in a solid in the presence of the magnetic field. It
corresponds to the generation of the elastic twist mandated by
conservation of energy and total angular momentum. This mode of
relaxation must be the dominant one for a rigid spin cluster
embedded in an elastic medium. Simple universal expression, Eqs.\
(\ref{MatrElLatRes}) and (\ref{MatrElLabRes}) with Eq.\ (\ref
{OmegatilDef}), for the transition matrix element has been derived
for an arbitrary spin Hamiltonian and an arbitrarily directed
magnetic field. The method of computing the transition matrix
elements, presented in this paper, consists of two steps. In the
first (needed by any theory) step one must obtain the eigenstates
of the spin Hamiltonian, the transition between which is going to
be studied. Once these eigenstates are known, the computation of
the spin-phonon transition rate by our method reduces to the
calculation of the matrix elements of $\mathbf{S}$ between the
eigenstates of interest. We have studied in detail the transitions
between tunnel-split spin states, which are difficult to compute
by conventional methods. The role of the time-reversal symmetry
has been analyzed. As demonstrated in Sec.\ \ref
{Sec-MatrElem-TR}, not all transitions between spin states of
different parity with respect to time reversal are ruled out by
time-reversal symmetry. This is at variance with the case of
spatial symmetries.

Universal formulas have been obtained for two particular cases of
the strong uniaxial anisotropy. The first case is that of the
tunnel splitting created by the crystal field, Eq.\ (\ref{W0Res}).
The second case if that of the tunnel splitting created by the
transverse magnetic field in the presence of an uniaxial magnetic
anisotropy, Eq.\ (\ref{W0KxKy}). The first case corresponds to,
e.g., the intensively studied $S=10$ Fe$_{8}$ molecular
cluster, while the second case corresponds to, e.g., the recently studied $%
S=4\;$ Ni$_{4}$ molecular cluster. \cite{barkenyanhen04prl} For both cases,
we have obtained relaxation rates, and their field dependence, in terms of
independently measurable constants and with no adjustable parameters. Among
other applications, these rates are responsible for the linewidths measured
in electron spin resonance experiments. Our results opens the way for an
accurate comparison between theory and experiment on the field dependence of
the relaxation rate. Among many possible applications our theory also has an
important consequence for the industry of spin-based solid-state qubits. It
provides parameter-free, mandated by symmetry, lower bound on the
decoherence of any such qubit.

While multiphonon processes can be studied by the same method, this paper is
limited to the rates of direct one-phonon processes that dominate
spin-lattice transitions at low temperature. Two-phonon Raman processes that
contribute at higher temperatures will be studied elsewhere.


\section{Acknowledgements}

The work of E.M.C. has been supported by the NSF Grant No. EIA-0310517. R.S.
gratefully acknowledges stimulating discussions with Martin Reuter.

\appendix

\section{Spin-phonon relaxation for adjacent levels}

\label{App-adjacent}

Here we shall study the spin Hamiltonian
\begin{equation}
\hat{H}_{S}=\hat{H}_{A}+\hat{H}_{Z}=-DS_{z}^{2}-g\mu _{B}H_{z}S_{z}.
\label{HAUNiaxial}
\end{equation}
The \emph{exact} energy levels of this Hamiltonian are given by
Eq.\ (\ref {EmDef}) with $E_{m}^{(A)}=-Dm^{2}.$ We consider the
spin-phonon relaxation between the adjacent levels of
$\hat{H}_{S},$ $m$ and $m^{\prime }=m+1.$ The energy difference
between these levels is
\begin{eqnarray}
W &=&\hbar \omega _{0}=E_{m}-E_{m+1}  \nonumber \\
&=&D(2m+1)+g\mu _{B}H_{z}.  \label{WUniaxial}
\end{eqnarray}
(we assume $W>0$). We first study the $(m,m+1)$ transitions by a
conventional method that employs tilting of the anisotropy axis by
transverse phonons, see, e.g., Refs.\
\onlinecite{harpolvil96,garchu97}. In this method one writes
$\hat{H}_{A}$ in the form
\begin{equation}
\hat{H}_{A}(\mathbf{n})=-D\left( \mathbf{n\cdot S}\right) ^{2},
\label{HATilted}
\end{equation}
where $\mathbf{n}$ is the direction of the anisotropy axis,
\begin{equation}
\mathbf{n=e}_{z}+\delta \mathbf{n},\qquad \delta \mathbf{n}=\delta \mathbf{%
\phi \times n.}  \label{nTilted}
\end{equation}
Expanding $\hat{H}_{A}(\mathbf{n})$ up to the linear terms in $\delta
\mathbf{\phi }$ one obtains $\hat{H}_{A}(\mathbf{n})\cong -DS_{z}^{2}+\hat{H}%
_{\mathrm{s-ph}}$ with
\begin{eqnarray}
&&\hat{H}_{\mathrm{s-ph}}=-D\left[ \left( \delta \mathbf{n\cdot S}\right)
\left( \mathbf{e}_{z}\mathbf{\cdot S}\right) +\left( \mathbf{e}_{z}\mathbf{%
\cdot S}\right) \left( \delta \mathbf{n\cdot S}\right) \right] =  \nonumber
\\
&&-D\left[ S_{x}S_{z}+S_{z}S_{x}\right] \delta \phi _{y}+D\left[
S_{y}S_{z}+S_{z}S_{y}\right] \delta \phi _{x}.  \label{HsphRot}
\end{eqnarray}
The same result follows from the calculation of the commutator in
Eq.\ (\ref {Smallphi}). We shall calculate the transition matrix
element between the states $\left| \Psi _{+}\right\rangle =\left|
m\right\rangle \otimes \left| n_{\mathbf{k}}\right\rangle $ and
$\left| \Psi _{-}\right\rangle =\left| m+1\right\rangle \otimes
\left| n_{\mathbf{k}}+1\right\rangle $. One obtains
\begin{equation}
\left\langle \Psi _{-}\left| \hat{H}_{\mathrm{s-ph}}\right| \Psi
_{+}\right\rangle =\frac{iD}{2}l_{m,m+1}\left( 2m+1\right) \langle n_{%
\mathbf{k}}+1|\delta \phi _{-}|n_{\mathbf{k}}\rangle ,  \label{Marteluniax}
\end{equation}
where $l_{m,m+1}=\sqrt{S(S+1)-m(m+1)}$ and
\begin{equation}
\delta \phi _{-}\equiv \delta \phi _{x}-i\delta \phi _{y}\,.
\label{chirality}
\end{equation}
Note that the contribution to Eq.\ (\ref{Marteluniax}) of the
transverse phonons of definite chirality, Eq.\ (\ref{chirality}),
having $l_{z}=1$ projection of the angular momentum onto the
$z$-axis, is in accordance with the conservation of the total
angular momentum for the ($m+1,m$)-transition.

Quantization of lattice rotations with the help of Eq.\
(\ref{deltaphiPh}) yields Eq.\ (\ref{MatrElwPhononsquant}) with
[cf. Eq.\ (\ref{VklamDef})]
\begin{equation}
V_{\mathbf{k}\lambda }=\frac{iD}{2}l_{m,m+1}\left( 2m+1\right) \frac{e^{i%
\mathbf{k\cdot r}}\left[ \mathbf{k}\times \mathbf{e}_{\mathbf{k}\lambda }%
\right] \cdot (\mathbf{e}_{x}-i\mathbf{e}_{y})}{\sqrt{8M\hbar \omega _{%
\mathbf{k}\lambda }}}.  \label{Phik}
\end{equation}
The relaxation rates between the states $m$ and $m+1$ are given by
Eqs.\ (\ref
{WpmRes}) and (\ref{WmpDef}) with $V_{\mathbf{k}\lambda }$ defined by Eq.\ (%
\ref{Phik}). With the help of the formulas listed in the Appendix \ref
{App-math} one obtains
\begin{equation}
W_{0}=\frac{(2m+1)^{2}l_{m,m+1}^{2}}{24\pi \hbar }\frac{D^{2}}{Mv_{t}^{2}}%
\frac{\omega _{m,m+1}^{3}}{\omega _{D}^{3}}.  \label{W0mmp1Res}
\end{equation}
The limit of $D\rightarrow 0$ corresponds to a free spin. In that
limit, although the states $\left| m\right\rangle $ and $\left|
m+1\right\rangle $ are still separated in energy due to the
magnetic field, the rotation of the lattice cannot cause any
relaxation and Eq.\ (\ref{W0mmp1Res}) produces zero result.

It is instructive to see how the above result can be obtained by our new
method using Eqs.\ (\ref{MatrElLabRes}) and (\ref{OmegaplusminusRes}) with ${%
\mathbf{\Omega }}_{-+}=\langle n_{\mathbf{k}}+1|{\mathbf{\hat{\Omega}}}|n_{%
\mathbf{k}}\rangle $. As the transition $\left| m\right\rangle \rightarrow $
$\left| m+1\right\rangle $ is due to the spin operator $S_{+}$, one obtains
\begin{equation}
\left\langle \Psi _{-}\left| \hat{H}_{\mathrm{s-ph}}\right| \Psi
_{+}\right\rangle =-\frac{\hbar }{2}l_{m,m+1}\langle n_{\mathbf{k}}+1|\hat{%
\Omega}_{-}|n_{\mathbf{k}}\rangle ,  \label{MatrElmmp1newmethod}
\end{equation}
where $\hat{\Omega}_{-}\equiv \hat{\Omega}_{x}-i\hat{\Omega}_{y}.$ For $%
\mathbf{H}=H_{z}\mathbf{e}_{z}$ this gives
\begin{eqnarray}
&&\left\langle \Psi _{-}\left| \hat{H}_{\mathrm{s-ph}}\right| \Psi
_{+}\right\rangle =  \nonumber \\
&&\frac{i}{2}l_{m,m+1}\left( \hbar \omega _{0}-g\mu _{B}H_{z}\right) \langle
n_{\mathbf{k}}+1|\delta \phi _{-}|n_{\mathbf{k}}\rangle .
\label{MatrElmmp1newmethod2}
\end{eqnarray}
Here the magnetic field cancels according to Eq.\
(\ref{WUniaxial}) and one obtains Eq.\ (\ref{Marteluniax}).

\section{Role of rotations around the $z$ axis}

\label{App-E}

In Sec.\ \ref{Sec-SPhRelax-anis} we have calculated the
spin-phonon relaxation rate between the tunnel-split resonance
states of the spin in the case when the tunnel splitting $\Delta $
is due to the terms in the crystal-field Hamiltonian $\hat{H}_{A}$
that do not commute with $S_{z}.$ The result, Eq.\ (\ref{W0Res}),
is universal and it does not depend on the detailed form of
$\hat{H}_{A}.$ We have seen that the relaxation is generated by
the phonons rotating the lattice around the $z$ axis. The physics
of this process is transparent: in the rotating frame the spin
feels the magnetic field along the $Z$-axis. This field couples to
$S_z$, which produces the relaxation. In this Appendix we will
obtain this result for a biaxial spin model by the conventional
method used in the previous Appendix. We shall demonstrate that in
the conventional method the relaxation arises from the rotation of
the hard axis by the phonons. As we shall see, the computation of
this effect, even for the simplest spin Hamiltonian, is
significantly more cumbersome that the computation of the
transition rate by our new method.

Consider the spin Hamiltonian $\hat{H}_{S}$ of Eq.\ (\ref{HSDef}) with $\hat{H%
}_{Z}$ given by Eq.\ (\ref{HZDef}) and
\begin{equation}
\hat{H}_{A}=-DS_{z}^{2}+E\left( S_{x}^{2}-S_{y}^{2}\right) =-DS_{z}^{2}+%
\frac{1}{2}E\left( S_{+}^{2}+S_{-}^{2}\right) .  \label{HAbiax}
\end{equation}
We assume $E>0$ for simplicity. If $E$ is negative, one can change
its sign by rotating the coordinate system by $\pi /2$ around the
$z$ axis. For $E\ll D$ the energy levels of this system can be
approximately characterized by the quantum number $m.$ Any two
levels $m^{\prime }>$ $m$ can be brought into resonance by the
longitudinal component of the magtetic field $H_{z}$, see Sec.\
\ref{Sec-MatrElem}. The tunnel splitting $\Delta $ can be
calculated with the help of the high-order perturbation theory:
\begin{equation}
\frac{\Delta }{2}=V_{m,m+2}\frac{1}{E_{m+2}-E_{m}}V_{m+2,m+4}\ldots
V_{m^{\prime }-2,m^{\prime }},  \label{DeltaEPert}
\end{equation}
where
\begin{equation}
V_{m,m+2}=\frac{1}{2}E\left\langle m\left| S_{-}^{2}\right| m+2\right\rangle
=\frac{1}{2}El_{m,m+1}l_{m+1,m+2}.  \label{Vmmplus2}
\end{equation}
Calculation in Eq.\ (\ref{DeltaEPert}) yields \cite{gar91jpa}
\begin{eqnarray}
\Delta &=&\frac{2D}{\left[ \left( m^{\prime }-m-2\right) !!\right] ^{2}}
\nonumber \\
&&\times \sqrt{\frac{\left( S+m^{\prime }\right) !\left( S-m\right) !}{%
\left( S-m^{\prime }\right) !\left( S+m\right) !}}\left( \frac{E}{2D}\right)
^{(m^{\prime }-m)/2}.  \label{DeltaERes}
\end{eqnarray}
One can see from this calculation that\ in our case\ in Eq.\ (\ref{TLS}) is $%
e^{i\varphi }=1.$ Let us now consider the phonons that rotate the lattice
around the $z$ axis. It is easy to obtain that the result of such rotation
is
\begin{equation}
\hat{H}_{\mathrm{s-ph}}=-iE\left( S_{+}^{2}-S_{-}^{2}\right) \delta \phi
_{z}\equiv \hat{V}\delta \phi _{z}.  \label{HsphE}
\end{equation}
Now, similarly to the calculation in Sec.\
\ref{Sec-MatrElem-beyond}, one can compute the matrix element of
$\hat{H}_{\mathrm{s-ph}}$ between slightly
delocalized states $\left| \tilde{m}\right\rangle $ and $\left| \tilde{m}%
^{\prime }\right\rangle $ [see Eq.\ (\ref{mmprimetildeDef})] by
building a minimal perturbative chain between the pure states
$\left| m\right\rangle $
and $\left| m^{\prime }\right\rangle $ and identifying the result with $%
\Delta $ given by Eq.\ (\ref{DeltaEPert}). As
$\hat{H}_{\mathrm{s-ph}}$ can be inserted at $\left( m^{\prime
}-m\right) /2$ positions in the chain, one obtains
\begin{equation}
\left\langle \tilde{m}\left| S_{-}^{2}\right| \tilde{m}^{\prime
}\right\rangle =\left\langle \tilde{m}^{\prime }\left| S_{+}^{2}\right|
\tilde{m}\right\rangle =\frac{m^{\prime }-m}{2}\frac{\Delta }{E}
\label{MS2pmS}
\end{equation}
and thus
\begin{eqnarray}
\left\langle \tilde{m}\left| \hat{V}\right| \tilde{m}^{\prime }\right\rangle
&=&i\frac{m^{\prime }-m}{2}\Delta ,  \nonumber \\
\left\langle \tilde{m}^{\prime }\left| \hat{V}\right| \tilde{m}\right\rangle
&=&-i\frac{m^{\prime }-m}{2}\Delta .  \label{MESph}
\end{eqnarray}
For the spin eigenfunctions given by Eq.\ (\ref{Psipm}) with $e^{i\varphi }=1$%
one obtains
\begin{equation}
\left\langle \psi _{-}\left| \hat{V}\right| \psi _{+}\right\rangle =i\frac{%
m^{\prime }-m}{2}\,\Delta .  \label{Vminusplus}
\end{equation}
The spin-phonon matrix element is then given by
\begin{equation}
\left\langle \Psi _{-}\left| \hat{H}_{\mathrm{s-ph}}\right| \Psi
_{+}\right\rangle =i\frac{m^{\prime }-m}{2}\,\Delta \,\delta \phi _{-+,z}.
\label{MatrElEconv}
\end{equation}
The latter coincides with the result obtained by our new method,
Eq.\ (\ref {MatrElwPhonons}) with $\left\langle \psi _{-}\left|
S_{z}\right| \psi _{+}\right\rangle $ given by Eq.\
(\ref{MatrElSz}) and $\hbar \Omega _{-+,z}=i\hbar \omega _{0}\phi
_{-+,z}$. The advantage of the new method is apparent as it only
involves the trivial computation of the matrix element of $S_{z}$.

To gain a deeper insight into the difference between the two
methods, one can start with the general form of the eigenstates
$\left| \Psi _{\pm }\right\rangle $ given by Eq.\ (\ref{PsikGen})
and compare general expressions for the spin-phonon matrix element
due to the rotation around the $z$ axis, used by the two methods.
The traditional method uses the worked out form of the commutator
in Eq.\ (\ref{Smallphi}) while the new method uses Eq.\ (
\ref{CommDisappeares}). The equivalence of the two results relies
on the identity
\begin{equation}
\sum_{m=-S}^{S}C_{+,m}^{\ast }V_{m,m+2}C_{-,m+2}=\frac{\Delta }{4}%
\sum_{m=-S}^{S}C_{+,m}^{\ast }mC_{-,m},  \label{SumRuleSplitting}
\end{equation}
where $\Delta =E_{+}-E_{-}$ and the matrix element $V_{m,m+2}$ is
due to the terms in $\hat{H}_{A}$ that do not commute with
$S_{z}.$ (We consider for illustration the model with
$V_{m,m+1}=0$ and zero bias, $H_{z}=0.$) The left-hand side of
Eq.\ (\ref{SumRuleSplitting}) corresponds to the worked out
commutator and it vanishes if one approximates $\left| \Psi _{\pm
}\right\rangle $ by the two-state model of Sec.\
\ref{Sec-MatrElem-twostate}, \ i.e., if one neglects all $C_{\pm
,m^{\prime \prime }}$ with $m^{\prime \prime }\neq m,m^{\prime }.$
Leaving $C_{\pm ,m}$ with all $m$ leads to the correct result of
order $\Delta $ in the left-hand side of Eq.\ (\ref
{SumRuleSplitting}), but only after one accurately accounts for
the cancellation of many small terms containing powers of $E/D\ll
1.$ On the contrary, in the right-hand side of Eq.\
(\ref{SumRuleSplitting}) the small quantity $\Delta $ has been
already factored out, and it is sufficient to make the two-state
approximation in the sum over $m.$ \newline

\section{Role of the time-reversal symmetry}

\label{App-TR}

In quantum mechanics time reversal is represented by the operator $\hat{K}$
that satisfies $\hat{K}^{-1}=\hat{K}^{\dagger },$ and, in addition, performs
complex congugation.\cite{mes76} The latter comes from the requirement that
the lhs, $i\hbar \partial _{t}\left| \psi (t)\right\rangle $ of the
time-dependent Schr\"{o}dinger equation is covariant under $t\rightarrow -t$
(which changes $i\rightarrow -i),$ only, if one takes its complex conjugate.
The latter makes $\hat{K}$ \emph{antilinear}:
\begin{equation}
\hat{K}\left[ c_{1}\left| \psi _{1}\right\rangle +c_{2}\left| \psi
_{2}\right\rangle \right] =c_{1}^{\ast }\hat{K}\left| \psi _{1}\right\rangle
+c_{2}^{\ast }\hat{K}\left| \psi _{2}\right\rangle .  \label{KAntilinear}
\end{equation}
Because of the antilinearity, one has to specify whether $\hat{K}$ acts to
the right (by default) or to the left:
\begin{equation}
\left\langle \psi \left| \hat{K}\right| \varphi \right\rangle \equiv
\left\langle \psi \left| \left( \hat{K}\right| \varphi \right\rangle \right)
=\left( \left\langle \psi \left| \hat{K}\right) \right| \varphi
\right\rangle ^{\ast }.  \label{Brackets}
\end{equation}
Operator $\hat{K}$ is called \emph{antiunitary} because of its antilinearity
and its property $\hat{K}^{-1}=\hat{K}^{\dagger }.$

Transformation of the spin operator $\mathbf{S}$ under time reversal follows
from the requirement that $\mathbf{S}$ should behave as the orbital momentum
$\mathbf{L}$ and it thus should change its sign under time reversal:
\begin{equation}
\mathbf{S}^{\prime }=\hat{K}\mathbf{S}\hat{K}^{\dagger }=\hat{K}\mathbf{S}%
\hat{K}^{-1}=-\mathbf{S}.  \label{eq3}
\end{equation}
An arbitrary spin state $\left| \psi \right\rangle $ can be represented as a
superposition of the eigenstates $\left| m\right\rangle $ of $S_{z}$.
Choosing the phase factor of $\left| m\right\rangle $ to be one, i.e.,
taking $\left| m\right\rangle $ as real makes \ the matrix elements $%
\left\langle m\left| S_{z}\right| m^{\prime }\right\rangle $ and $%
\left\langle m\left| S_{\pm }\right| m^{\prime }\right\rangle $ real. In
that case the explicit form of the time-reversal operator is $\hat{K}=Q%
\hat{U}=\hat{U}Q,$ where $\hat{U}=\exp \left( -i\pi S_{y}\right) $ is a
unitary operator and $Q$ makes complex conjugation. Indeed, $\hat{U}$
changes the signs of $S_{x}$ and $S_{z},$ whereas $Q$ changes the sign of $%
S_{y},$ so that Eq.\ (\ref{eq3}) is fulfilled.

In order to determine the action of $\hat{K}$ on the tunnel-split spin
eigenstates $\left| \psi _{\pm }\right\rangle $ of the crystal-field
Hamiltonian $\hat{H}_{A}$ we have at first to determine the action of $\hat{K%
}$ on $\left| m\right\rangle $. Application of
$\hat{K}S_{z}=-S_{z}\hat{K}$ and $\hat{K}S_{+}=-S_{-}\hat{K}$
[that follow from Eqs.\ (\ref{eq3}) and (\ref {KAntilinear})] onto
$\left| m\right\rangle $ yields, up to an irrelevant global
factor,

\begin{equation}
\hat{K}|m\rangle =(-1)^{S-m}|-m\rangle .  \label{eq5}
\end{equation}
As a consequence one obtains $\hat{K}^{2}|m\rangle =(-1)^{2S}\left|
m\right\rangle ,$ i.e., $\hat{K}^{2}=1$ for integer spins $S.$

In zero field $\hat{H}_{S}\ $reduces to $\hat{H}_{A}$ that must be
time-reversible, $\hat{K}\hat{H}_{A}\hat{K}^{-1}=\hat{H}_{A},$ i.e., it
satisfies $\left[ \hat{K},\hat{H}_{A}\right] =0.$ For integer $S$ the
tunnel-split eigenstates $|\psi _{\pm }\rangle $ of $\hat{H}_{A}$ are
nondegenerate, thus they must be eigenstates of $\hat{K}$:
\begin{equation}
\hat{K}\left| \psi _{\pm }\right\rangle =\varepsilon _{\pm }\left| \psi
_{\pm }\right\rangle   \label{eq11}
\end{equation}
with eigenvalues $\varepsilon _{\pm }.$ Since $\hat{K}^{2}=1$ for integer
spin $S,$ it must be $\varepsilon _{\pm }\in \{-1,1\}$. Let us prove now
that $\varepsilon _{+}$ and $\varepsilon _{-}$ have different signs. To this
end, we decompose $\hat{H}_{A}$ into the longitudinal and transverse parts, $%
\hat{H}_{A}^{\mathrm{long}}$ \ and $\hat{H}_{A}^{\mathrm{trans}}.$ The
former satisfies $\left[ S_{z},\hat{H}_{A}^{\mathrm{long}}\right] =0$ and
thus it has $|\pm m\rangle $ as at least twofold degenerate eigenstates. The
transverse part $\hat{H}_{A}^{\mathrm{trans}}$ has matrix elements between
different $|m\rangle $ and thus it removes the degeneracy between $|\pm
m\rangle $ for integer $S$. Let us introduce

\begin{equation}
\hat{H}_{A}(\lambda )=\hat{H}_{A}^{\mathrm{long}}+\lambda \hat{H}_{A}^{%
\mathrm{trans}}  \label{eq10}
\end{equation}
with $\lambda \,\,$\thinspace r{eal. }In the limit $\lambda \rightarrow 0$
one can find $\left| \psi _{\pm }\right\rangle $ analytically. Using Eqs.\ (%
\ref{eq5}) and (\ref{KAntilinear}), as well as the relation $(-1)^{2m}=1$
for integer $S$, one can check that

\begin{equation}
\left| \psi _{\pm }(0)\right\rangle =\frac{1}{\sqrt{2}}\left[ e^{i\chi
}|m\rangle \pm e^{-i\chi }|-m\rangle \right]   \label{eq12}
\end{equation}
satisfy Eq.\ (\ref{eq11}) with
\begin{equation}
\varepsilon _{\pm }(0)=\pm (-1)^{S-m}.  \label{eq13}
\end{equation}
Since $\hat{H}_{A}(\lambda )$ is continuous in $\lambda ,$ the eigenstates $%
|\psi _{\pm }(\lambda )\rangle $ and, in turn $\varepsilon _{\pm }(\lambda ),
$ must be continuous, too. This continuity and the discreteness of $%
\varepsilon _{\pm }$ implies that $\varepsilon _{\pm }$ are independent of $%
\lambda .$ Thus one obtains $\varepsilon _{+}(\lambda )=-\varepsilon
_{-}(\lambda )$ for all $\lambda ,$ including $\lambda =1$. Thus $|\psi
_{\pm }\rangle \equiv |\psi _{\pm }(1)\rangle $ inherits its parity from the
unperturbed eigenstates. Consequently, we get an interesting result that the
tunnel split eigenstates of an \emph{arbitrary} crystal field Hamiltonian
must have \emph{opposite} parity with respect to time reversal symmetry,
i.e.,
\begin{equation}
\varepsilon _{+}\varepsilon _{-}=-1.  \label{epsilonproduct}
\end{equation}

The spin-phonon Hamiltonian $\hat{H}_{\mathrm{s-ph}}$ is invariant under
time reversal, i.e., $\ \hat{H}_{\mathrm{s-ph}}^{\prime }=\hat{K}\hat{H}_{%
\mathrm{s-ph}}\hat{K}^{\dagger }=\hat{H}_{\mathrm{s-ph}}.$ One can obtain a
symmetry relation for the spin matrix element if one makes time reversal of
both $\hat{H}_{\mathrm{s-ph}}$ and the spin states. Inserting $\hat{K}%
^{\dagger }K=\hat{K}^{-1}\hat{K}=1$ into the matrix element one proceeds as
follows:

\begin{eqnarray}
&&\left\langle \psi _{-}\left| \hat{H}_{\mathrm{s-ph}}\right| \psi
_{+}\right\rangle  \nonumber \\
&=&\left\langle \psi _{-}\left| \hat{K}^{\dagger }K\hat{H}_{\mathrm{s-ph}}%
\hat{K}^{\dagger }K\right| \psi _{+}\right\rangle  \nonumber \\
&=&\left\langle \psi _{-}\left| \hat{K}^{\dagger }\hat{H}_{\mathrm{s-ph}}%
\hat{K}\right| \psi _{+}\right\rangle  \nonumber \\
&=&\left[ \left( \left\langle \psi _{-}\right| \hat{K}^{\dagger }\right)
\hat{H}_{\mathrm{s-ph}}\hat{K}\left| \psi _{+}\right\rangle \right] ^{\ast }
\nonumber \\
&=&\varepsilon _{+}\varepsilon _{-}\left\langle \psi _{-}\left| \hat{H}_{%
\mathrm{s-ph}}\right| \psi _{+}\right\rangle ^{\ast }  \nonumber \\
&=&-\left\langle \psi _{-}\left| \hat{H}_{\mathrm{s-ph}}\right| \psi
_{+}\right\rangle ^{\ast },  \label{HsphTrans}
\end{eqnarray}
where we have used Eqs.\ (\ref{Brackets}), (\ref{eq11}), and (\ref
{epsilonproduct})$.$ This proves Eq.\ (\ref{eq1}). Spin operator
changes sign
under time reversal, Eq.\ (\ref{eq3}), thus a similar procedure leads to Eq.\ (%
\ref{STRev}).

\section{Useful relations}

\label{App-math}

To calculate the spin-phonon relaxation rates, the following useful
relations can be used.

Summation over the two transverse polarizations:
\begin{equation}
\left[ \mathbf{k}\times \mathbf{e}_{\mathbf{k}t_{1}}\right] =\pm k\mathbf{e}%
_{\mathbf{k}t_{2}}  \label{AnotherTrans}
\end{equation}
and
\begin{equation}
\sum_{t=t_{1},t_{2}}\left( \mathbf{e}_{\mathbf{k}t}\cdot \mathbf{a}\right)
\left( \mathbf{e}_{\mathbf{k}t}\cdot \mathbf{b}\right) =\left( \mathbf{%
a\cdot b}\right) -\frac{\left( \mathbf{k\cdot a}\right) \left( \mathbf{%
k\cdot b}\right) }{k^{2}}.  \label{transverseraus}
\end{equation}

Averaging over the directions of the vector $\mathbf{k}$:
\begin{equation}
\left\langle \left( \mathbf{k}\cdot \mathbf{a}\right) \left( \mathbf{k}\cdot
\mathbf{b}\right) \right\rangle =\frac{k^{2}}{3}\left( \mathbf{a}\cdot
\mathbf{b}\right) .  \label{kAver}
\end{equation}

\bibliographystyle{prsty}
\bibliography{gar-oldworks,gar-books,gar-own,gar-tunneling,gar-relaxation}

\begin{thebibliography}{10}

\bibitem{wal32}
{I. Waller}, Z. Phys. {\bf 79},  370  (1932).

\bibitem{heitel34}
{W. Heitler and E. Teller}, Proc. R. Soc. London A {\bf 155},  629  (1934).

\bibitem{kro39}
{R. de L. Kronig}, Physica (Amsterdam) {\bf 6},  33  (1939).

\bibitem{vle40}
{J. H. Van Vleck}, Phys. Rev. {\bf 57},  426  (1940).

\bibitem{orb61}
{R. Orbach}, Proc. R. Soc. London A {\bf 264},  458  (1961).

\bibitem{abrble70}
{A. Abragam and A. Bleaney}, {\em Electron {P}aramagnetic {R}esonance of
  {T}ransition {I}ons} (Clarendon Press, Oxford, 1970).

\bibitem{calcal65pr}
{E. Callen and H. B. Callen}, Phys. Rev. {\bf 139},  A455  (1965).

\bibitem{mel72prl}
{R. L. Melcher}, Phys. Rev. Lett. {\bf 28},  165  (1972).

\bibitem{bonmel76prb}
{L. Bonsall and R. L. Melcher}, Phys. Rev. B {\bf 14},  1128  (1976).

\bibitem{mel79prb}
{R. L. Melcher}, Phys. Rev. B {\bf 19},  284  (1979).

\bibitem{fed75prb}
{P. A. Fedders}, Phys. Rev. B {\bf 12},  2045  (1975).

\bibitem{fedmel76prb}
{P. A. Fedders and R. L. Melcher}, Phys. Rev. B {\bf 14},  1142  (1976).

\bibitem{fed77prb}
{P. A. Fedders}, Phys. Rev. B {\bf 15},  3297  (1977).

\bibitem{dohful75}
{V. Dohm and P. Fulde}, Z. Phys. B {\bf 21},  369  (1975).

\bibitem{wanlue77prb}
{P. S. Wang and B. L\"uthi}, Phys. Rev. B {\bf 14},  2972  (1977).

\bibitem{garchu97}
{D. A. Garanin and E. M. Chudnovsky}, Phys. Rev. B {\bf 56},  11 102  (1997).

\bibitem{chu94prl}
{E. M. Chudnovsky}, Phys. Rev. Lett. {\bf 72},  3433  (1994).

\bibitem{chumar02prb}
{E. M. Chudnovsky and X. Mart\'\i nes Hidalgo}, Phys. Rev. B {\bf 66},  054412
  (2002).

\bibitem{chu04prl}
{E. M. Chudnovsky}, Phys. Rev. Lett. {\bf 92},  120405  (2004).

\bibitem{chugar04prl}
{E. M. Chudnovsky and D. A. Garanin}, Phys. Rev. Lett. {\bf 93},  257205
  (2004).

\bibitem{mes76}
{A. Messiah}, {\em Quantum {M}echanics}, Vol.~II of {\em Lecture Notes in
  Physics} (Wiley, New York, 1976).

\bibitem{gar91jpa}
{D. A. Garanin}, J. Phys. A {\bf 24},  L61  (1991).

\bibitem{barkenyanhen04prl}
{E. del Barco, A. D. Kent, E. C. Yang, and D. N. Hendrickson}, Phys. Rev. Lett.
  {\bf 93},  157202  (2004).

\bibitem{harpolvil96}
{F. Hartmann-Boutron, P. Politi, and J. Villain}, Int. J. Mod. Phys. B {\bf
  10},  2577  (1996).

\end{thebibliography}

\end{document}